\documentclass{PoS}

\usepackage{epsfig}
\def\text{{}}   \newcommand{\tr}{{\rm Tr}}
\def\lsim{\raise0.3ex\hbox{$<$\kern-0.75em\raise-1.1ex\hbox{$\sim$}}}
\def\gsim{\raise0.3ex\hbox{$>$\kern-0.75em\raise-1.1ex\hbox{$\sim$}}}
\def\fm{\mathrm{fm}}

\title{Screening at finite temperature and density}

\ShortTitle{Screening at finite temperature and density}

\author{\speaker{Olaf Kaczmarek}\\
   Fakult\"{a}t f\"{u}r Physik,
 Universit\"{a}t Bielefeld, D-33615 Bielefeld, Germany\\
        E-mail: \email{okacz@physik.uni-bielefeld.de}
}

\date{\today}

\abstract{
We present lattice QCD results on heavy quark free energies, extract from its
temperature dependence entropy and internal energy contributions, and discuss
the onset of medium effects that lead to screening of static quark-antiquark
sources in a thermal medium.
Most results are obtained in (2+1)-flavour QCD on a line of
constant physics with almost realistic quark masses and compared to previous
results from 2-flavor QCD as well as pure gauge theory. 
Furthermore, we discuss results on the density dependence of screening masses
that have been obtained using a leading order Taylor expansion in the baryon
chemical potential.
}

\FullConference{Critical Point and Onset of Deconfinement
          4th International Workshop\\
		 July 9-13 2007\\
		 GSI Darmstadt,Germany}

\begin{document}
\section{Introduction}
In-medium properties of heavy quarks and heavy quark bound states are of
fundamental interest for the understanding of strongly interacting matter at
high temperatures and densities probed in current and future heavy ion
collision experiments at RHIC, FAIR and LHC.
The temperature dependence of heavy quark free energies, potential energies as
well as
screening masses and radii are important ingredients for studies on strongly
coupled quark gluon plasma, possible existence of heavy quark bound states in
the QGP as well as on transport properties of heavy quarks in the high
temperature phase.\\
We present preliminary results on heavy quark free energies \cite{rbcbi_hq}
that are based on an analysis of gauge field configurations generated by the
RBC-Bielefeld collaboration in (2+1)-flavor QCD for the calculation of the QCD
equation of state \cite{Cheng:2006qk}.
The pion mass is about 220 MeV and the strange quark mass is adjusted to its
physical value.
The calculations were performed with improved staggered fermions on lattices
with temporal extent $N_\tau=4$ and 6 accompanied by high statistics zero
temperature calculations to set the scale and to extract the zero temperature
potential and corresponding renormalization constants.\\
We perform a renormalization of the finite temperature heavy quark free
energies and of the Polyakov loop using the renormalization constants obtained at
zero temperature.
We will discuss the relation between heavy quark free energies,
entropy and internal energy contributions and analyze their critical behavior
in the transition region.
The temperature dependence of the interaction between static quark
anti-quark pairs will be analyzed in terms of, in general distance and
temperature dependent, running couplings and screening masses.
A preparatory study of the density dependence of screening masses in 2-flavor
QCD with large quark masses indicates that non-perturbative effects in the
behavior of screening is dominated by the gluonic sector.
\section{Zero temperature potential}
\begin{figure}[t]
\center\epsfig{file=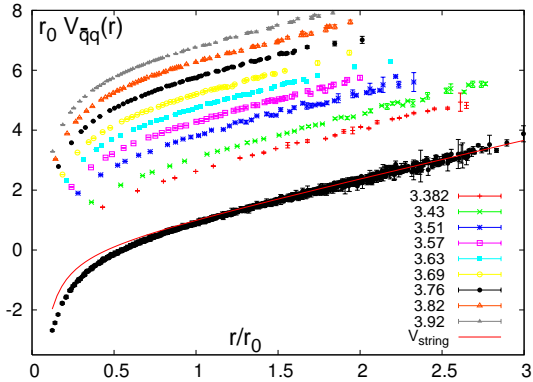,width=7.5cm}
\caption{
Un-renormalized zero temperature potentials for different values of $\beta$ and
renormalized overall $T=0$-potential (lower black points) in units of $r_0$. The
solid line shows the string potential as discussed in the text.
}
\label{zerot.fig}
\end{figure}
\begin{figure}[t]
  \epsfig{file=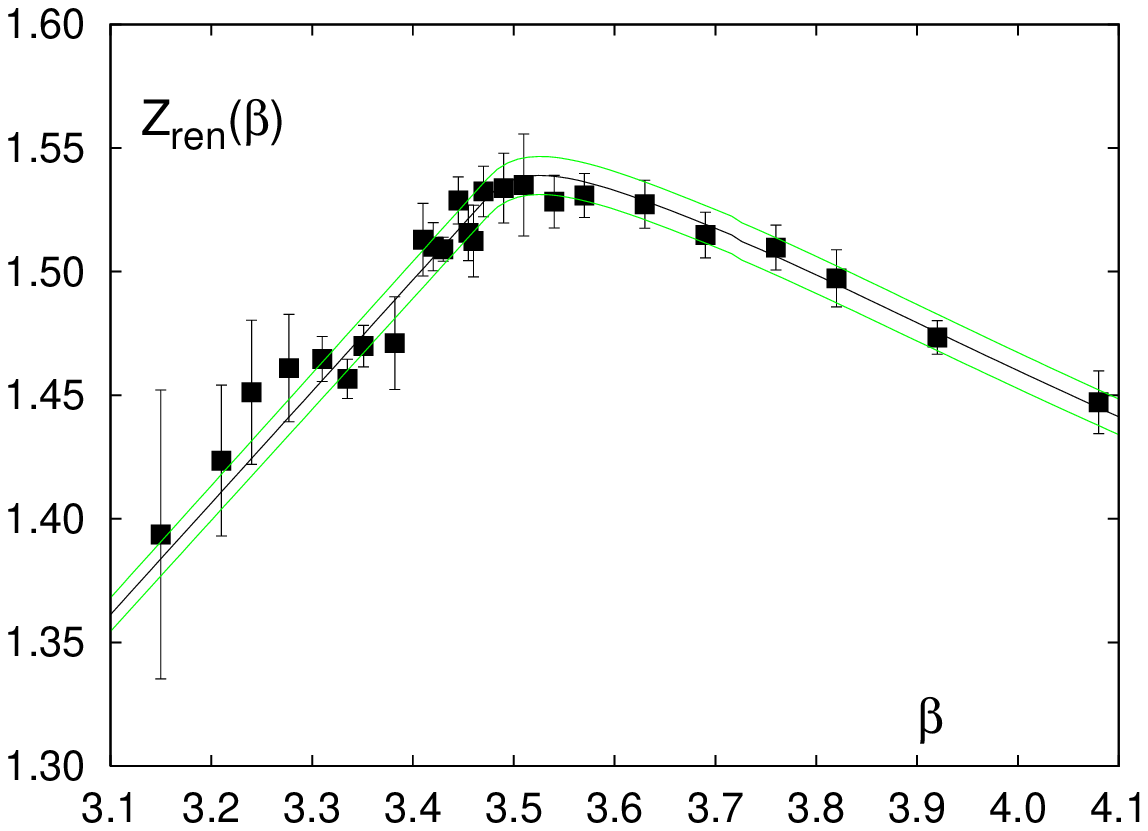,width=7.5cm}
  \epsfig{file=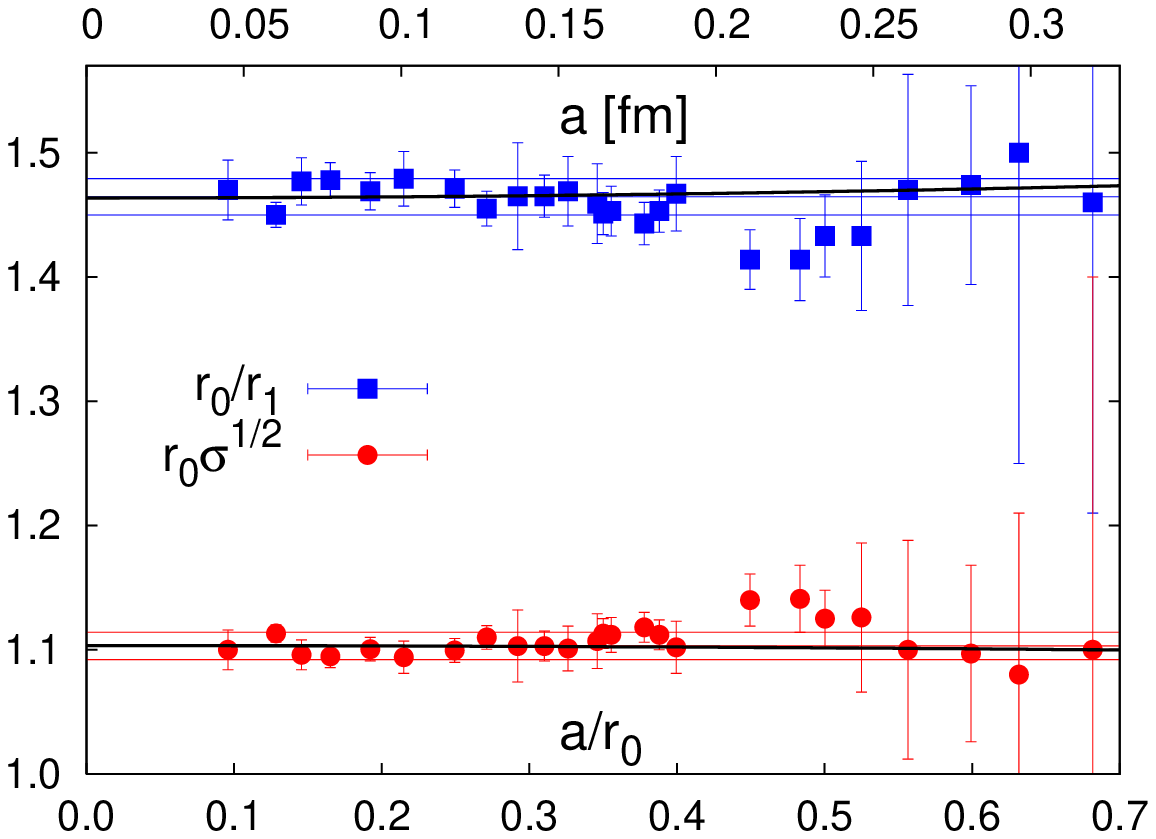,width=7.5cm}
\caption{
Renormalization constants, $Z_{\mathrm{ren}}(\beta)$, obtained from the matching of the
zero temperature potentials (left). Dimensionless combinations, $r_0/r_1$ and
$r_0 \sqrt{\sigma}$. The lower axis denotes the scale $(r_0/a)^{-1}$ and the
upper axis the lattice cut-off in physical units. The lines are explained in
the text (right).
}
\label{r0r1.fig}
\end{figure}

In fig.~\ref{zerot.fig} we show the zero temperature potential,
$V_{\bar{q}q}(r)$, in units of the distance scale $r_0$ for various values of
the
coupling $\beta$ corresponding to different values of the lattice cut-off
varying from $a\simeq 0.3~\fm$ down to $a\simeq 0.05~\fm$. The scale $r_0/a$ is
define by the slope of the potential,
\begin{eqnarray}
\left( r^2\frac{d V_{\bar{q}q}(r)}{dr}\right)_{r=r_0}=1.65,
\label{r0.eq}
\end{eqnarray}
and can be used to convert to physical scales with a value $r_0=0.469(7)~\mathrm{fm}$
\cite{Gray:2005ur}. Additional parameters obtained from the slope of $V_{\bar{q}q}(r)$
are the string tension, $\sigma a^2$, and $r_1/a$ defined by replacing 1.65 by
1.0 in (\ref{r0.eq}). The dimensionless combinations of these parameters 
displayed in fig.~\ref{r0r1.fig} show only small cut-off effect,
e.g. $r_0\sqrt{\sigma}$ stays constant in the entire range of couplings in
which the lattice spacing changes by a factor of 6. Using a quadratic fit
Ansatz for $a\leq 0.15\fm$ we obtain $r_0\sqrt{\sigma}=1.1034(40)$ and
$r_0/r_1=1.4636(60)$.
For both ratios we observe that ${\cal O}(a^2)$ corrections are small.\\
The potentials as calculated on the lattice are ultraviolet divergent and need
to be renormalized. We have matched all potentials to a common value at large
distances, $r/r_0=1.5$, taken to be identical to the large distance string
potential which in units of $r_0$ is given by
\begin{eqnarray}
r_0 V_{\mathrm{string}}(r/r_0) = -\frac{\pi}{12 r/r_0} + (\sigma r_0^2)\frac{r}{r_0},
\end{eqnarray}
where we have used the value $r_0\sqrt{\sigma}$ quoted above. The
result of this renormalization is shown in the lower part of
fig.~\ref{zerot.fig}. The good matching of all potential data over the whole
distance range again shows that cut-off effects are small in this observable.\\
The matching procedure provides renormalization constants for the static quark
potential which we will use for the renormalization of Polyakov loops and heavy
quark free energies at finite temperatures. They are defined as
$Z_{\mathrm{ren}}(\beta)=\exp((c(\beta) a)/2)$, where $c(\beta)$ denotes the
constant shift for the corresponding zero temperature potential.
Already note here that no additional divergencies arise when going from zero to
finite temperatures. The solid line in fig.~\ref{zerot.fig} shows the string
potential. Note that the deviations of the potential $V_{\bar{q}q}(r)$ from
$V_{\mathrm string}(r)$ at small distances shows the effect of the QCD running
coupling while the Coulombic term in the string potential stems from large
distance corrections in the string model. Furthermore we do not observe string
breaking here which is due the operator used in our analysis and the limited
distance range analyzed here.
\section{Heavy quark free energies}
\begin{figure}[t]
  \epsfig{file=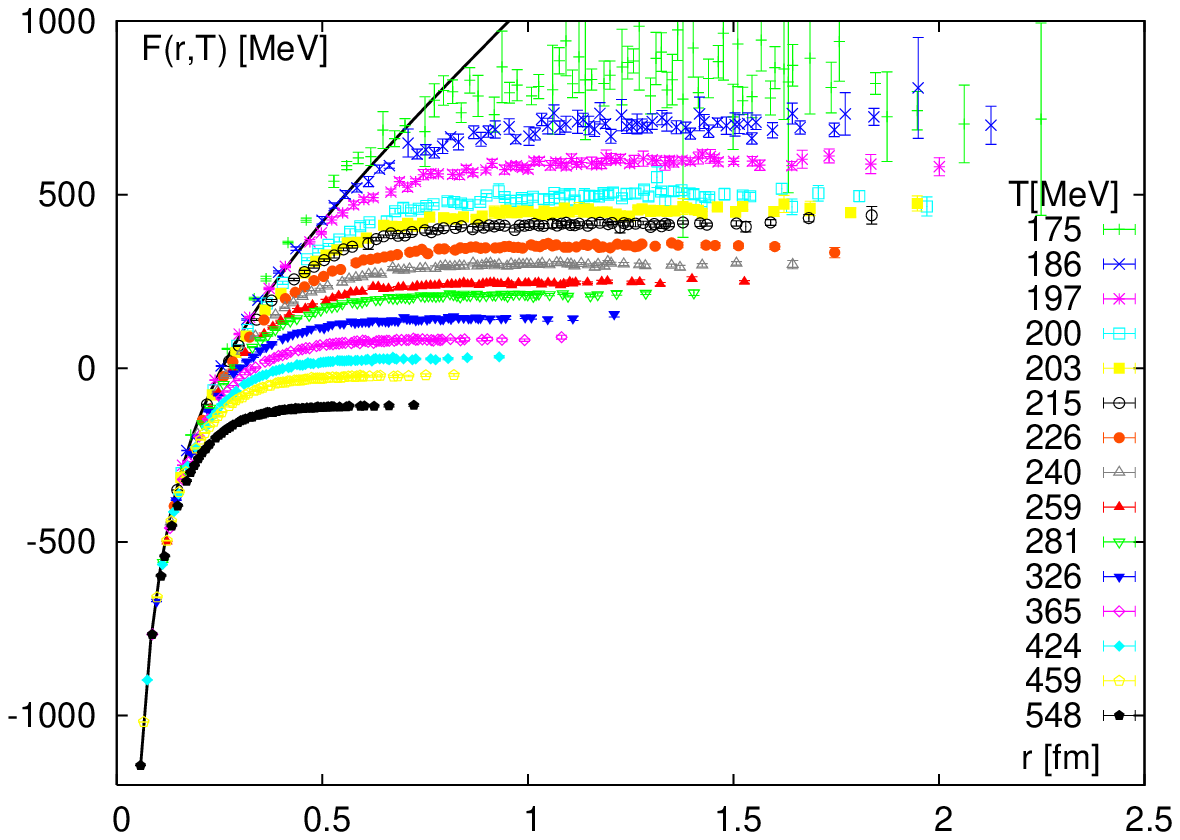,width=7.5cm}
  \epsfig{file=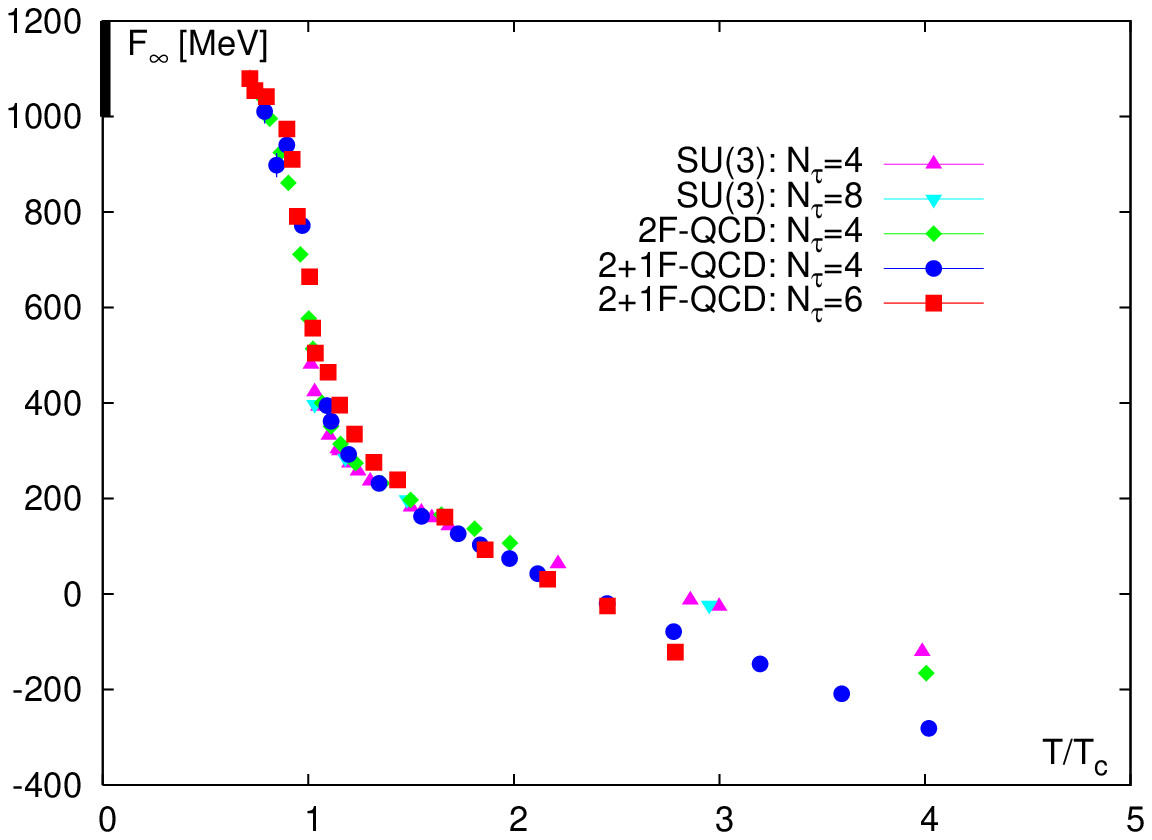,width=7.5cm}
\caption{
Renormalized heavy quark free energies, $F_1(r,T)$, for different values of the
temperature (left). The solid line shows the zero temperature potential
$V_1(r)$. Asymptotic values for the heavy quark free energies (right),
$F_1(r=\infty,T)$, for (2+1) flavors compared to earlier results for pure gauge
theory and 2-flavor \cite{Kaczmarek:2003ph,Kaczmarek:2005ui,Kaczmarek:2005gi}.
}
\label{hq.fig}
\end{figure}
The heavy quark free energy of a static quark-antiquark pair in a color
singlet state separated by distance $r$ is defined by
\begin{eqnarray}
\frac{F_1(\vec r,T)}{T} = -\log\left( \tr \left(L_{\mathrm{ren}}(\vec 0) L_{\mathrm{ren}}(\vec r)\right) \right),
\label{f1.eq}
\end{eqnarray}
where $L_{\mathrm{ren}}$ is the renormalized Polyakov loop,
\begin{eqnarray}
L_{\mathrm{ren}}(\vec x) = \left( Z_{\mathrm{ren}}(\beta)\right)^{N_\tau} L_{\mathrm{bare}} =
\prod_{x_0=0}^{N_\tau -1} Z_{\mathrm{ren}}(\beta) U_{(\vec x,x_0),0}\;.
\label{Lren.eq}
\end{eqnarray}
The operator used in (\ref{f1.eq}) in general is gauge dependent and we fix to
Coulomb gauge according to arguments based on \cite{Philipsen:2002az,Jahn:2004qr}.
As already noted we are using the renormalization constants, $Z_{\mathrm{ren}}(\beta)$,
obtained at zero temperature for the renormalization of the Polyakov loop. In
this way also the heavy quark free energies are properly renormalized, which is
evident from fig.~\ref{hq.fig}~(left) where $F_1(r,T)$ for different temperatures is
shown. The solid line is the zero temperature potential $V_{\bar{q}q}(r)$.\\
The free energies become temperature independent at small separations and
coincide with $V_{\bar{q}q}(r)$ showing the correct renormalization.
At larger distances temperature effects set in, below $T_c$ due to string
breaking and above $T_c$ due to screening of the static sources in the thermal
deconfined medium. The onset of this temperature effects is shifted towards smaller
distances with increasing temperature. 
The asymptotic values, $F_1(r=\infty,T)$, are show in fig.~\ref{hq.fig}~(right)
compared to previous results from quenched and 2-flavor QCD with
larger quark masses \cite{Kaczmarek:2003ph,Kaczmarek:2005ui,Kaczmarek:2005gi}. 
The qualitative behavior is comparable in
all those theories. Note that $F_\infty$ is infinite for the quenched
theory below the critical temperature. With dynamical quarks below the critical
temperature $F_\infty$ is
already close to the value estimated from zero temperature (indicated by the
black band), it shows a strong decrease around $T_c$ and a quite linear
behavior at high temperatures. This temperature dependence already indicates
that entropy contributions play an important role especially around the
transition temperature.
\section{Renormalized Polyakov loop}

As already discussed in the previous section, the renormalization constants
obtained at zero temperature can be used for the renormalization of the
Polyakov loop at finite temperature according to (\ref{Lren.eq}). Due to the
cluster property this renormalization is equivalent to a definition of
$L_{\mathrm{ren}}$ using the asymptotic value of the heavy quark free energy,
$F_\infty(T)=F_1(r=\infty,T)$, i.e. $L_{\mathrm{ren}}=\exp(-F(r=\infty,T)/2T)$.
While $L_{\mathrm{ren}}$ is zero below the critical temperature in pure SU(3) gauge
theory and has a finite gap at $T_c$ due to the first order transition in this
theory, it is non-zero in QCD with dynamical quarks even in the low temperature
phase. Although the temperature dependence is continuous here, $L_{\mathrm{ren}}$ shows
a pronounced rise around the transition region from small to large values in
the high temperature phase.\\
\begin{figure}[t]
  \epsfig{file=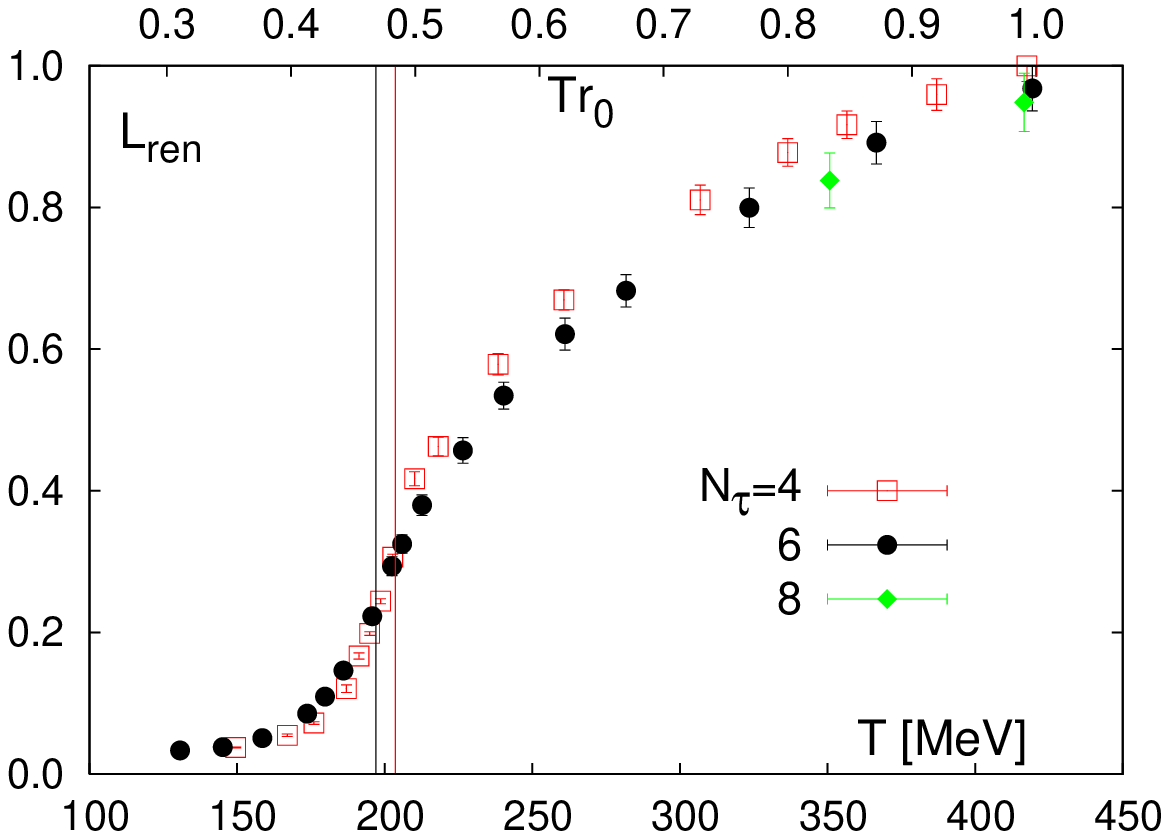,width=7.5cm}
  \epsfig{file=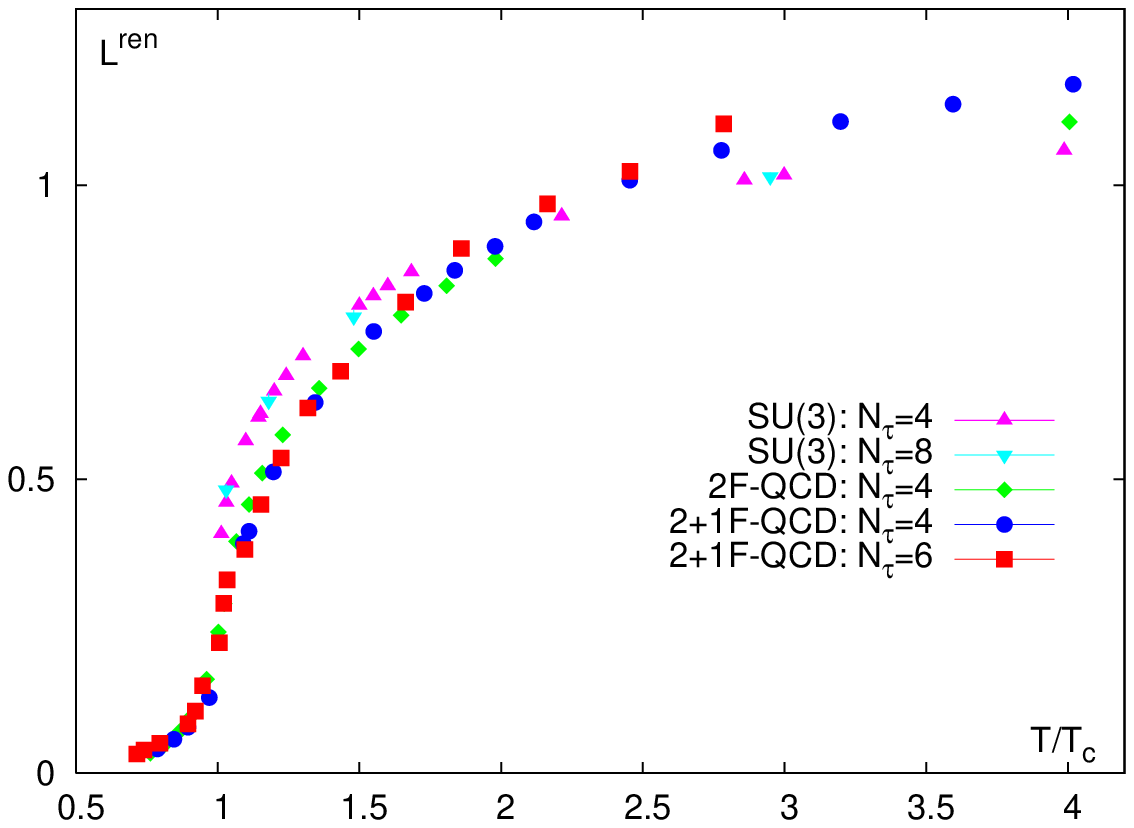,width=7.5cm}
\caption{
The renormalized Polyakov loop for different temporal lattice extents,
$N_\tau=4, 6$ and 8 (left). The vertical lines show the location of the transition
temperature determined in \cite{Cheng:2006qk} on lattices with temporal extent
$N_\tau=4$ (right line) and $N_\tau=6$ (left line), respectively. 
Comparison to earlier results (right) for pure gauge
theory and 2-flavor \cite{Kaczmarek:2003ph,Kaczmarek:2005ui,Kaczmarek:2005gi}.
}
\end{figure}
We note that the most rapid change is in good agreement with the region where
the chiral condensate as well as bulk thermodynamic quantities, e.g. energy and
entropy densities change most rapidly \cite{Cheng:2006qk}. Furthermore the cut-off
dependence of $L_{\mathrm{ren}}$ on lattices with temporal extent $N_\tau=4$ and 6 is
small, which is in agreement with results obtained in studies of $L_{\mathrm{ren}}$ in
pure SU(3) gauge theories \cite{Kaczmarek:2005ui,Kaczmarek:2002mc}. The large cut-off dependence observed
in a study with the 1-link stout smeared staggered action used in \cite{Aoki:2006br}
mainly seems to arise from the cut-off dependence of the zero temperature
observable ($f_K$) used to set the temperature scale. The fact that the
renormalized Polyakov loop becomes larger than one at high temperatures was
already observed in the quenched theory and is predicted by perturbation theory
where the high temperature limit, $L_{\mathrm{ren}}(T=\infty)=1$, is reached from
above \cite{Gava:1981qd,Petreczky:2005bd}. Note that the renormalized Polyakov loop as defined in (\ref{Lren.eq})
is no longer a SU(3) matrix.

\section{Entropy and internal energy contributions}
\begin{figure}[t]
  \epsfig{file=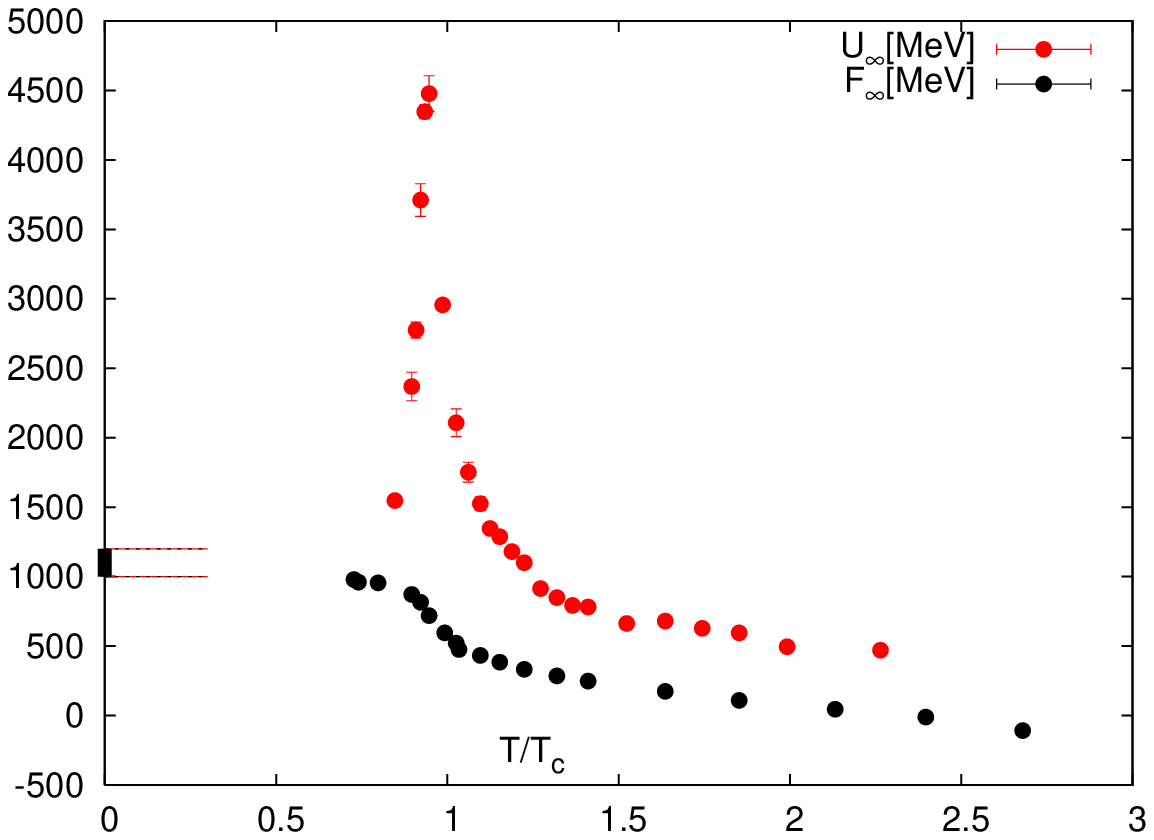,width=7.5cm}
  \epsfig{file=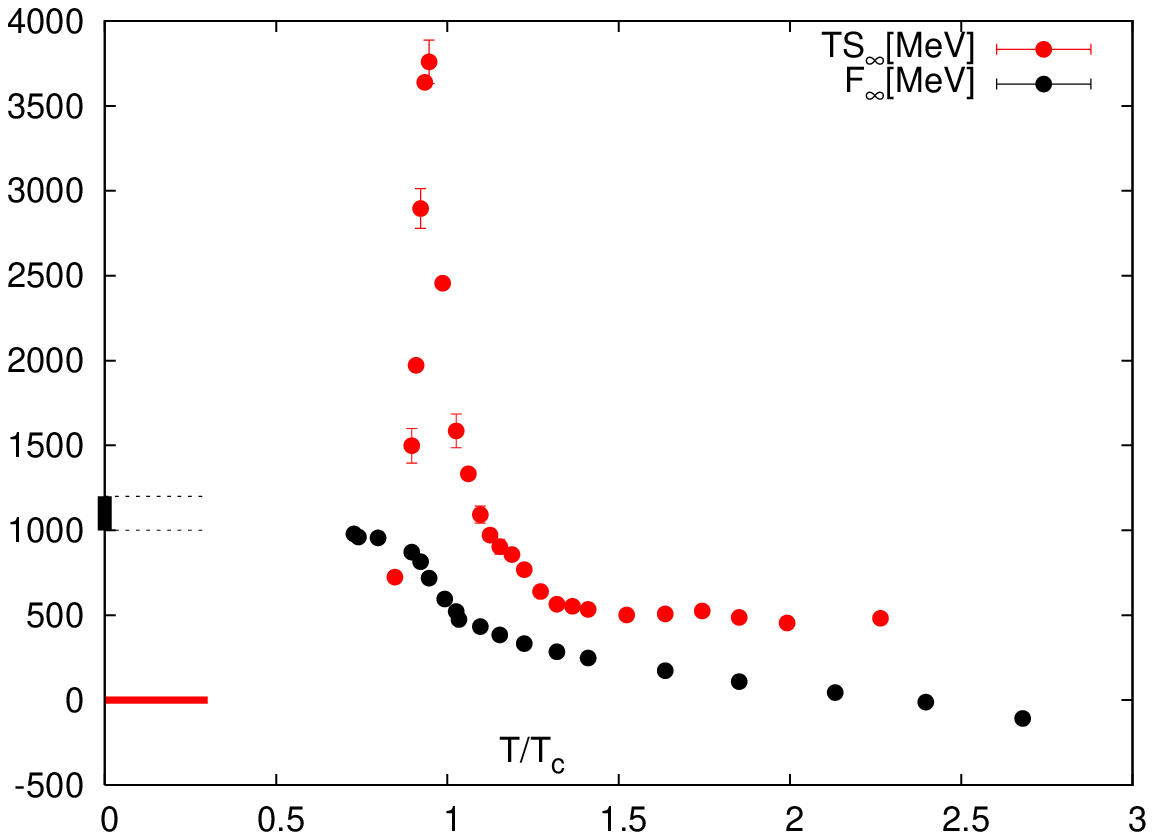,width=7.5cm}
\caption{
Asymptotic large distance values for internal energy (left) and entropy (right)
compared to heavy quark free energies. The left band indicates the zero
temperature limit of the energy and the red line shows the vanishing $T=0$
limit for $T S_\infty$.
}
\label{us.fig}
\end{figure}

The temperature dependence of the heavy quark-antiquark free energy at
asymptotic large distances, $F_\infty(T)$, already indicated that entropy
contributions play an important role especially in the vicinity of the
transition region. Thus it becomes important for discussions on strongly
coupled quark gluon plasma (SQGP)
\cite{Liao:2007mj,Shuryak:2003xe,Gyulassy:2004zy}, 
the possible existence of
heavy quark bound states that might survive above deconfinement 
\cite{Digal:2001iu,Mocsy:2005qw,Cabrera:2006wh,Park:2007qs,Alberico:2007rg,Shuryak:2004tx} 
as well as on
transport properties \cite{van Hees:2007me} in the high temperature phase.\\
Here we will only discuss the asymptotic large distance behavior of these
quantities, i.e. $S_\infty(T)=S_1(r=\infty,T)$ and $U_\infty(T)=U_1(r=\infty,T)$.
To separate entropy, $S_\infty$, and internal energy, $U_\infty$, contributions
from heavy quark free energies we use standard thermodynamic relations,
\begin{eqnarray}
S_\infty(T) = - \frac{\partial F_\infty(T)}{\partial T} \ \ \ \ \mathrm{and} \ \
\ \
U_\infty(T) = - \frac{\partial F_\infty(T)/T}{\partial T},
\end{eqnarray}
to calculated their asymptotic behavior. In fig.~\ref{us.fig} we compare
internal energy (left) and entropy (right) to the asymptotic behavior of the
free energy. The band at the left axis indicates the value of the energy in the
zero temperature limit and the line in the right figure indicates that entropy,
$T S_\infty$, vanishes at zero temperature.\\
In contrast to the monotonic decrease observed for $F_\infty$, both $U_\infty$
and $T S_\infty$ show qualitative different (critical) behavior in the
transition region. While at small temperatures they rapidly approach values
close to their zero temperature limits, both show a pronounced peak,
$U_\infty(T\simeq T_c)\simeq 4.5~\mathrm{GeV}$ and
$T S_\infty(T\simeq T_c)\simeq 4~\mathrm{GeV}$, which again decreases rapidly
toward smaller values and become rather flat above $1.5~T_c$. Note
that the temperature where both observables attain their maximum is in good
agreement with the critical temperature obtained using different observables in
\cite{Cheng:2006qk}.
Similar behavior was already observed in previous studies with $n_f$=2 and $3$
with larger quark masses in \cite{Kaczmarek:2005gi,Petreczky:2004pz}.
\section{Effective running coupling constant}
\begin{figure}[t]
  \epsfig{file=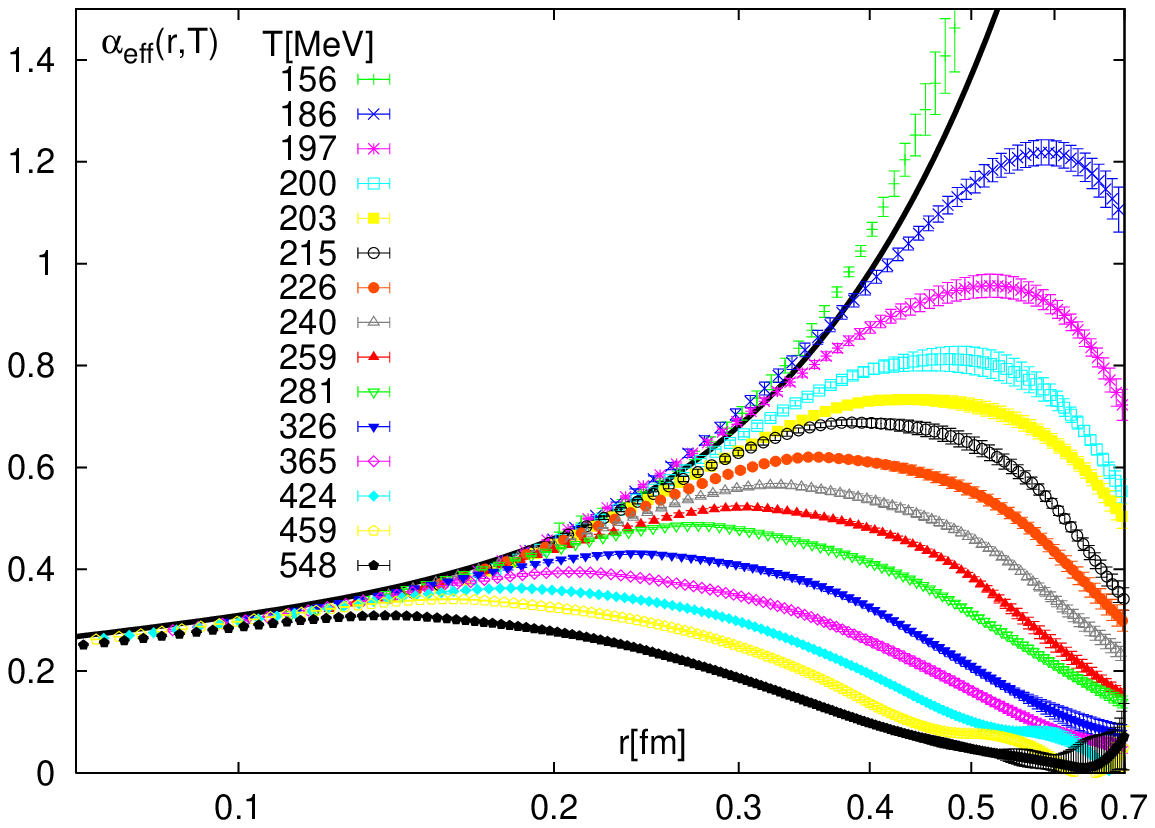,width=7.5cm}
  \epsfig{file=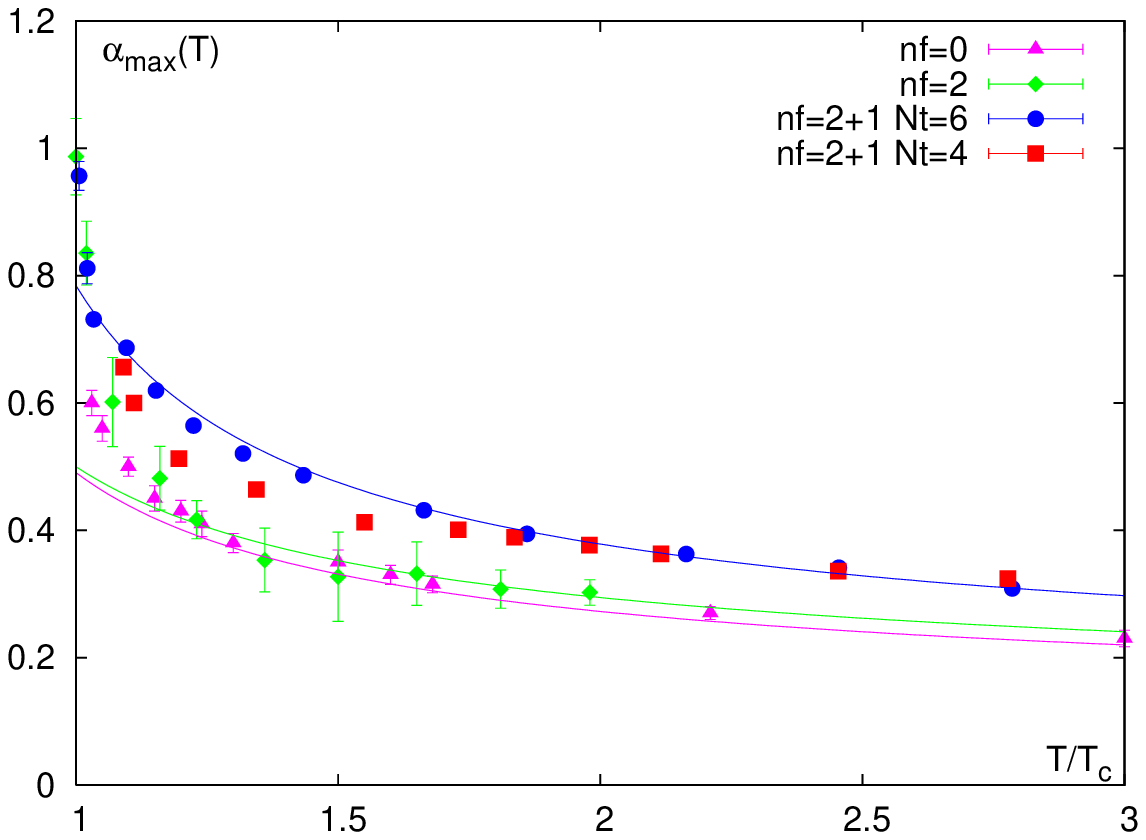,width=7.5cm}
\caption{
The effective (distance dependent) running coupling (left). The solid line
shows the zero temperature running coupling calculated from the zero
temperature potential. Effective temperature dependent coupling (right),
$\alpha_{\mathrm{max}}(T)$, define by the maximum of $\alpha_{\mathrm{eff}}(r,T)$ as explained in
the text for (2+1) flavor compared to 2-flavor and pure gauge
theory \cite{Kaczmarek:2003ph,Kaczmarek:2005ui,Kaczmarek:2005gi}. The lines
show a perturbative inspired fit-Ansatz for high temperatures.
}
\label{coupl.fig}
\end{figure}

As already discussed, the heavy quark-antiquark free energies become
temperature independent at small separations. To analyze the onset of
temperature effects it is more appropriate to study the effective distance and
temperature dependent running coupling defined through
\begin{eqnarray}
\alpha_{\mathrm{eff}}(r,T)&\equiv&\frac{3}{4}r^2\frac{dF_1(r,T)}{dr}\;.
\end{eqnarray}
In fig.~\ref{coupl.fig}~(left) results for various temperatures are compared to
the zero temperature running coupling, $\alpha_{T=0}(r)$, defined
in the same way (solid line). Note that the quadratic rise of $\alpha_{T=0}$ is
a non-perturbative effect that stems from the
linear rising string tension term in the potential, while at small distances
the logarithmic weakening of the coupling is visible and at sufficiently small
distances it should reach the perturbative (asymptotic free) behavior.\\
At finite temperatures, $\alpha_{\mathrm{eff}}(r,T)$ follows this zero temperature
behavior to relatively large distances, before screening sets in leading to a
maximum and a decrease at larger distances. Although the onset of temperature
effects and the maximum is shifted towards smaller distances with increasing
temperature, at temperatures slightly above the critical one,
$\alpha_{\mathrm{eff}}(r,T)$ still follows the quadratic behavior quite far, indicating
that even above $T_c$ remnants of confinement forces are present.\\
We have used the value of the maximum of the running coupling,
$\alpha_{\mathrm{max}}(T)$,
to define an effective temperature dependent coupling constant that can be used
to indicate an effective coupling strength at distances where the screening of
the static quark-antiquark pair sets in. The results are shown in
fig.~\ref{coupl.fig}~(right) in comparison with pure SU(3) gauge theory and
2-flavor QCD results from \cite{Kaczmarek:2005ui}.
The solid lines show results of a fit with an Ansatz using the two-loop
perturbative coupling at temperatures $T\geq 1.2T_c$.
Note that the 2-flavor results are calculated for rather large quark masses
($m/T=0.4$). Therefore the increase of $\alpha_{\mathrm{max}}$ for (2+1)-flavor is rather
a quark mass effect than a flavor effect. 
The comparison of the results for temporal extent $N_\tau=4$ and 6 again shows
that cut-off effects also in this observable are quite small.
The large value of the coupling close to the critical temperature should not be
identified with a large Coulombic coupling but clearly includes non-perturbative
effects.
\section{Debye screening at finite temperature and density}
\begin{figure}[t]
  \epsfig{file=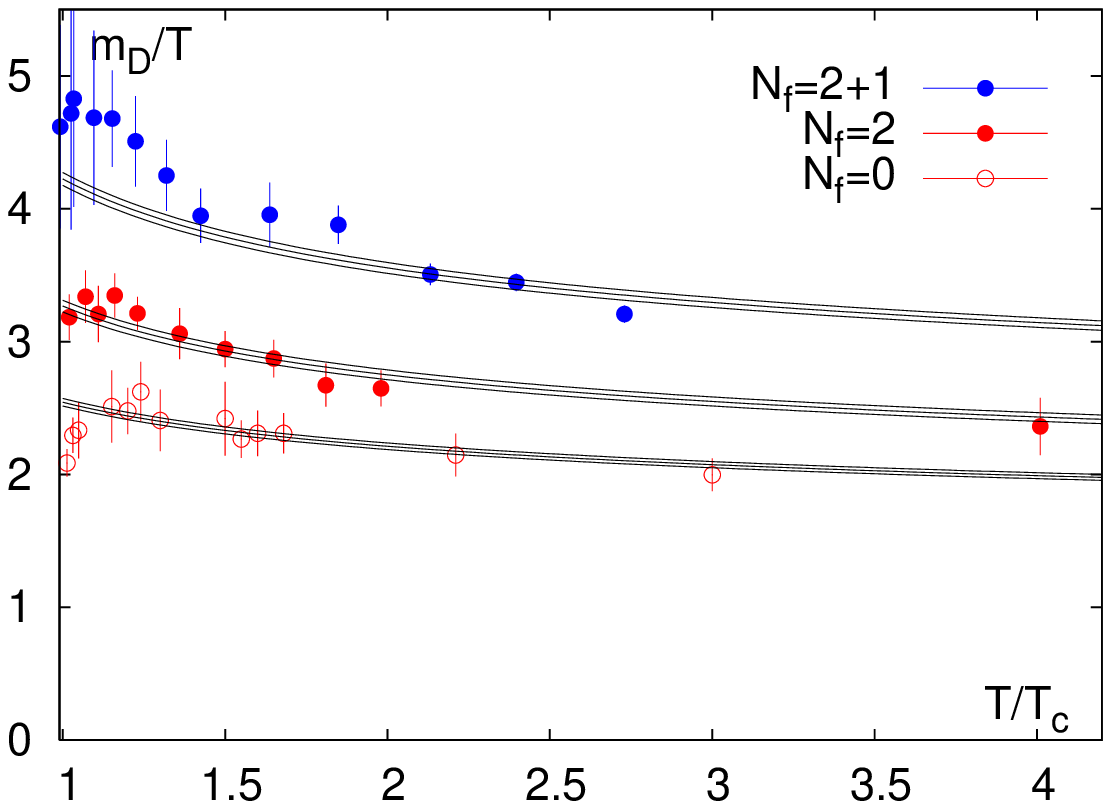,width=7.5cm}
  \epsfig{file=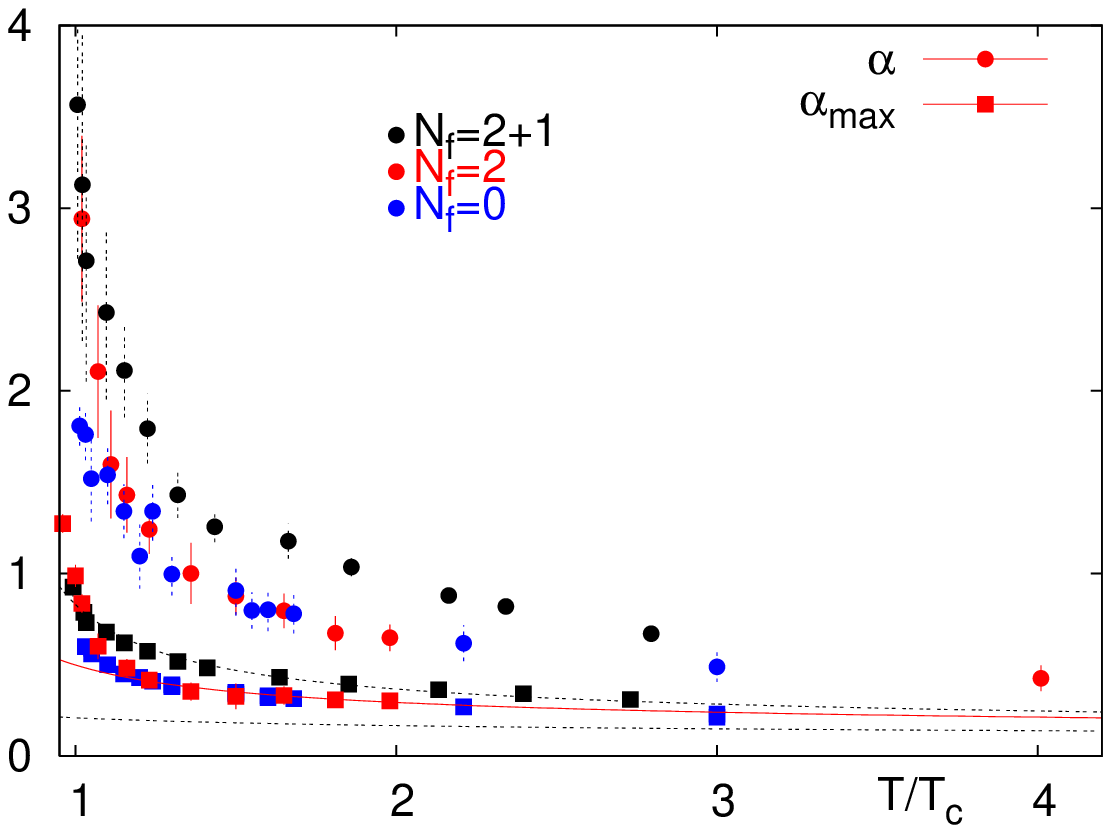,width=7.5cm}
\caption{
Debye screening masses in units of the temperature (left) for different number of
flavors. The solid lines and error band show the fit of a perturbative Ansatz
including a scale factor $A$ as described in the text. Effective T-dependent
Coulombic couplings (right) as obtained from the same fits compared to
$\alpha_{\mathrm max}$.
}
  \label{screen.fig}
\end{figure}

\begin{figure}[t]
  \epsfig{file=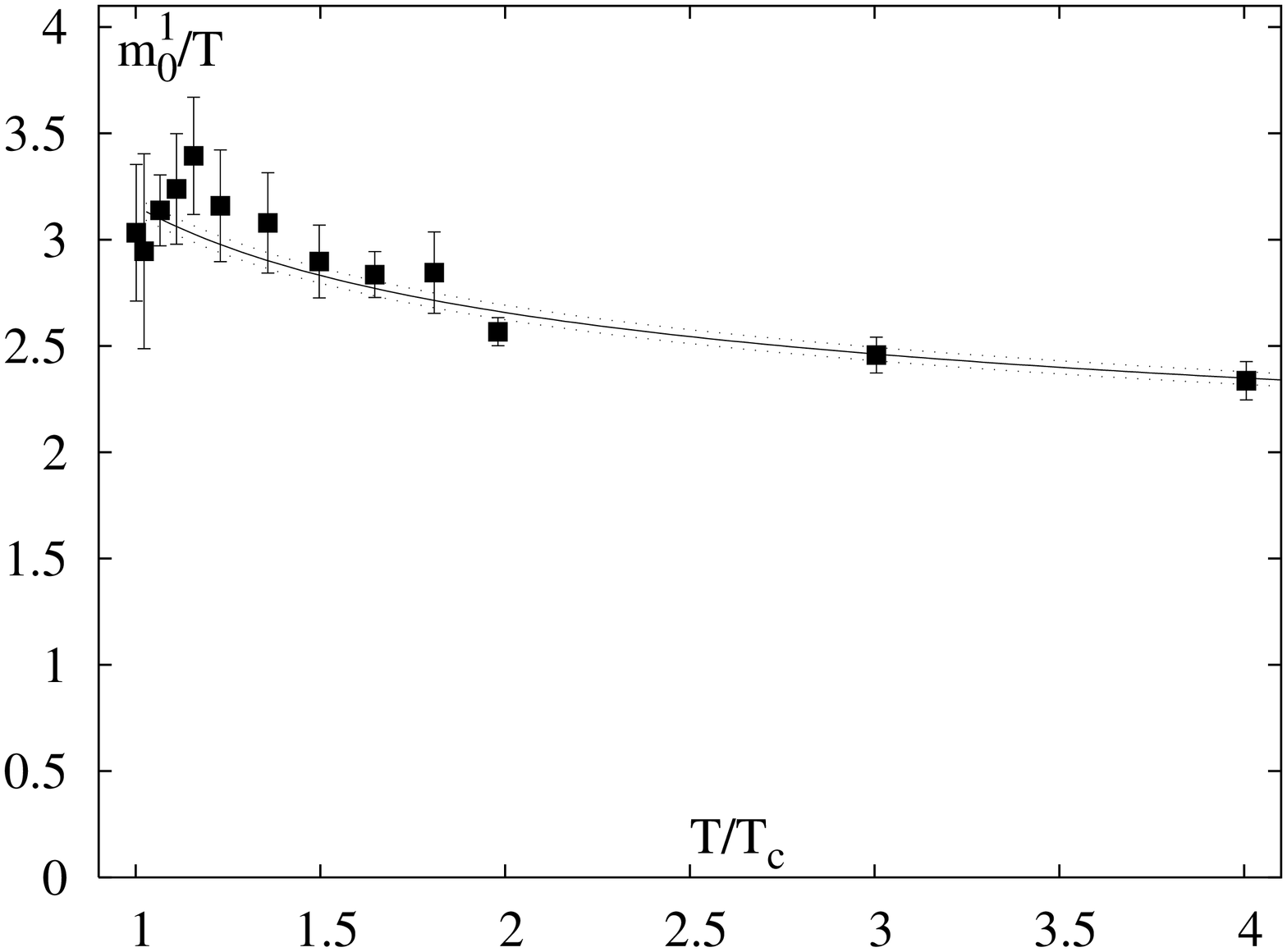,angle=270,width=7.5cm}
  \epsfig{file=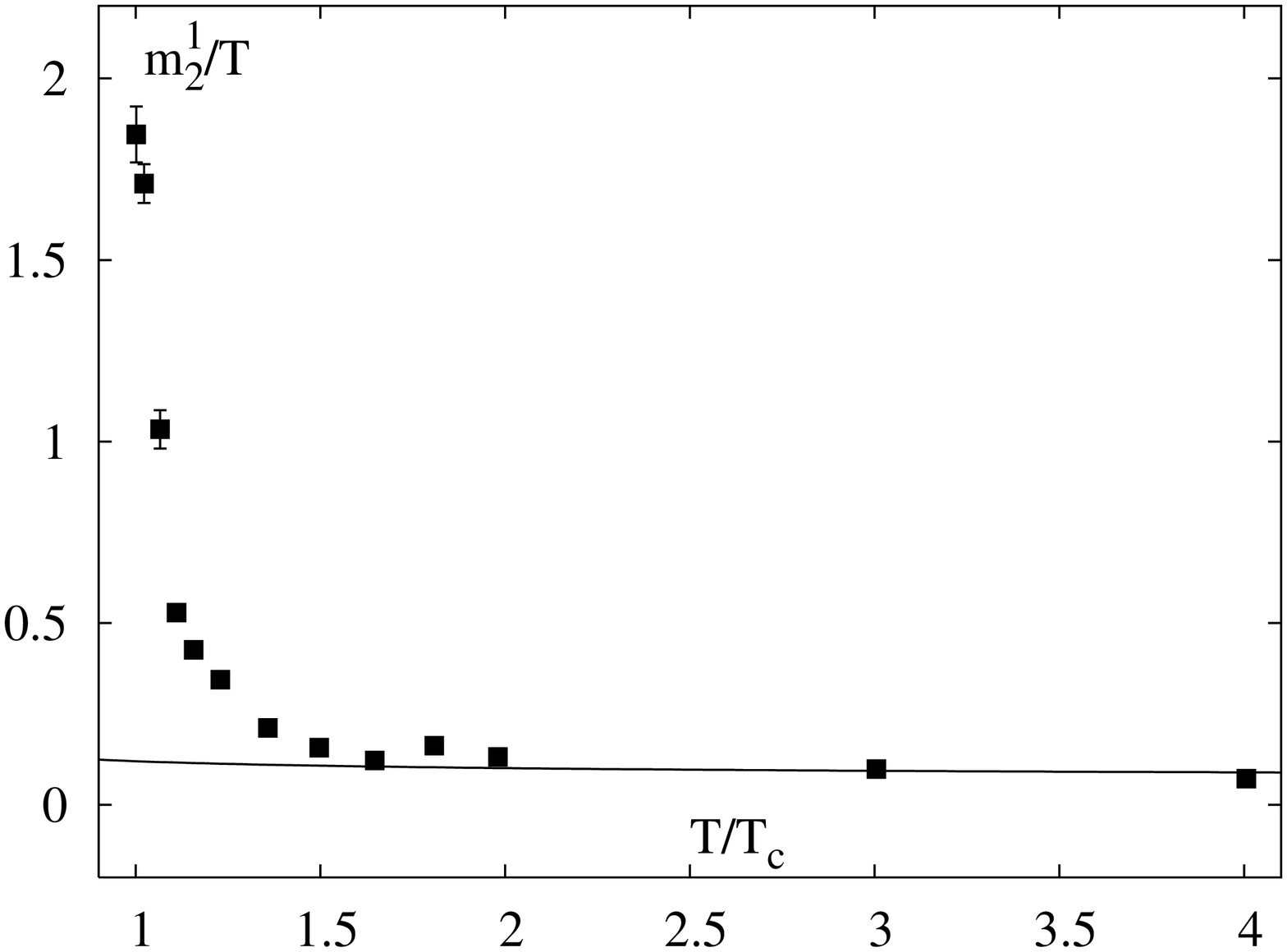,angle=270,width=7.5cm}
\caption{
Zero (left) and second (right) order density coefficients of the Taylor
expansion of heavy quark free energies in 2-flavor QCD \cite{Doring:2005ih}. The
solid line in the left figure shows the fit and in the right figure the leading
order perturbative result as explained in the text.
}
\label{doering.fig}
\end{figure}

We follow the commonly used approach to define the non-perturbative (Debye)
screening mass, $m_D(T)$, and temperature dependent coupling constant,
$\alpha(T)$, using a screened Coulomb fit Ansatz for the large distance part of
the heavy quark free energies normalized by its asymptotic large distance
value,
\begin{eqnarray}
F_1(r,T)-F_1(r=\infty,T) = -\frac{4}{3}\frac{\alpha(T)}{r} e^{-m_D(T) r}.
\end{eqnarray}
Our results for the screening masses are summarized in
fig.~\ref{screen.fig}~(left) as function of $T/T_c$ and are compared to results
in pure gauge theory and 2-flavor QCD from \cite{Kaczmarek:2005ui}.\\
Although we are not expecting perturbation theory to hold at the small
temperatures analyzed here, the enhancement for increasing number of flavors is
in qualitative agreement with leading order perturbation theory, i.e.
\begin{eqnarray}
\frac{m_D(T)}{T} = A \left(1+\frac{N_f}{6}\right)^{1/2} g(T).
\end{eqnarray}
Here we have already introduced a multiplicative constant $A$ to allow for
non-perturbative contributions ($A=1$ in perturbation theory). Using a two-loop
perturbative definition of $g(T)$ a best fit analysis for $m_D(T)/T$ for
temperatures $T\geq 1.2 T_c$ leads to $A=1.52(2)$, $1,42(2)$ and $1.66(2)$ for
$N_F=0$, $2$ and $(2+1)$, respectively. The results including an error band are
shown by the solid lines in fig.~\ref{screen.fig}~(left).\\
The large value of $A$ indicates that non-perturbative contributions are
important in the temperature range analyzed here. Note that a study in pure
gauge theory up to temperatures as high as $24 T_c$ led to an only slightly
smaller value, $A=1.39(2)$, revealing that even at such high temperatures the
screening mass still is far from being perturbative.\\
The results for the coupling, $\alpha(T)$, are shown in
fig.~\ref{screen.fig}~(right) in comparison with $\alpha_{\mathrm{max}}$ as discussed in
the previous section. The large values of $\alpha(T)$, especially close to the
transition region, describes the large distance behavior of heavy quark free
energies and should not be confused with the coupling $\alpha_{\mathrm{max}}$ that
characterizes the short distance part of $F_1(r,T)$. The latter is almost
temperature independent and can to some extent be described by the zero
temperature coupling.\medskip\\
So far we have only discussed results for vanishing baryon number
densities. In leading order perturbation theory, the dependence of the Debye
mass on temperature and quark chemical potential is given by
\begin{eqnarray}
\frac{m_D(T,\mu_q)}{T} = g(T)\sqrt{1+ \frac{N_f}{6} 
+ \frac{N_f}{2\pi^2}\left( \frac{\mu_q}{T}\right)^2}.
\label{mupt.eq}
\end{eqnarray}
Here we will discuss results for two-flavor QCD with large quark masses \cite{Doring:2005ih}.
The extension to non-zero densities in terms of a Taylor expansion in the quark
chemical potential is straightforward 
and comparable to the Taylor expansion for bulk
thermodynamic observables \cite{Allton:2005gk},
\begin{eqnarray}
m_D(T,\mu_q) = m_0(T) + m_2(T) \left( \frac{\mu_q}{T}\right)^2 + {\cal O}(\mu_q^4),
\end{eqnarray}
where $m_0(T)$ is the zero-order (zero density) contribution as already
discussed above and $m_2(T)$ is the second order Taylor coefficient. Note that
for a $\bar{q}q$-system the odd order coefficients vanish.\\
The results for $m_0(T)$ and $m_2(T)$ are shown in
fig.~\ref{doering.fig}~(left) and (right), respectively. The solid line in the
left figure shows the fit result for $m_0$ as discussed above and the line in
the right figure shows the second order expansion coefficient from the
perturbative leading order expression (\ref{mupt.eq}). While the zero order
coefficient clearly shows non-perturbative behavior up to high temperatures,
the good agreement of the second order coefficient with the perturbative
expectation at temperatures above $1.5 T_c$ indicates that non-perturbative
contributions to screening in the high temperature phase of QCD are
dominated by the gluonic sector.\\
The density dependence of screening masses so far was only
analyzed for 2-flavor QCD and large quark masses. For a more realistic
description the analysis will be carried out for (2+1)-flavor QCD with almost
realistic quark masses in the future.

\section{Conclusions}
We have analyzed heavy quark free energies, their temperature dependence and
screening properties for (2+1)-flavor QCD with almost
realistic quark masses. The renormalization of the free energies as well as the
Polyakov loop was performed using renormalization constants obtained at zero
temperatures. For the asymptotic large distance behavior, entropy and internal
energy contributions were separated from the free energies and show critical
behavior around the transition region.\\
From the analysis of the effective running coupling constants the onset of
medium effects on a heavy quark-antiquark pair was analyzed and from the
large distance behavior of the free energies we extracted screening properties
at vanishing density. The extension to finite baryon densities so far is
limited to the case of 2-flavor QCD and large quarks masses. The agreement of
the second order expansion coefficient with perturbation theory at temperatures
above $1.5 T_c$ indicates that the main non-perturbative contributions to 
screening stems from the gluonic sector while fermionic contributions seem to
be small. This behavior has to be confirmed for (2+1)-flavor QCD with almost
realistic quark masses in the future.

\end{document}